# Direct visualization of antiferroelectric switching dynamics via electrocaloric imaging

Pablo Vales-Castro[1*], Miquel Vellvehi[2], Xavier Perpiñà[2], J.M.Caicedo[1], Xavier Jordà[2], Romain Faye[3], Krystian Roleder[4], Dariusz Kajewski[4], Amador Perez-Tomas[1], Emmanuel Defay[3], Gustau Catalan[1,5*]

[1] Catalan Institute of Nanoscience and Nanotechnology (ICN2), Campus Universitat Autonoma de Barcelona, Av. Serragalliners s/n, Bellaterra 08193, Spain; email: pablo.vales@icn2.cat

[2] Institut de Microelectrònica de Barcelona (IMB-CNM,CSIC), Carrer dels Til.lers s/n, Campus Universitat Autònoma de Barcelona (UAB), Cerdanyola del Vallès, 08193, Spain

[3] Materials Research and Technology Department, Luxembourg Institute of Science and Technology, University of Luxembourg, 2 avenue de l'Université, L-4365 Esch-sur-Alzette, Luxembourg

[4] Institute of Physics, University of Silesia in Katowice, ul. Uniwersytecka 4, 40-00 Katowice, Poland

[5] Institut Català de Recerca i Estudis Avançats (ICREA), Barcelona 08010, Catalunya; email: gustau.catalan@icn2.cat

**Abstract**

The large electrocaloric coupling in $PbZrO_3$ allows using high-speed infrared imaging to visualize antiferroelectric switching dynamics via the associated temperature change. We find that in ceramic samples of homogeneous temperature and thickness, switching is nucleation-limited and fast, with devices responding in the milisecond range. By introducing gradients of thickness, however, it is possible to change the dynamics from nucleation-limited to propagation-limited, whereby a single phase boundary sweeps across the sample like a cold front, at a speed of c.a. 20 cm/s. Additionally, introducing thermostatic temperature differences between two sides of the sample enables the simultaneous generation of a negative electrocaloric effect on one side and a positive one on the other, yielding a Janus-like electrocaloric response.

The electrocaloric effect (ECE) consists of a transient temperature change (ΔT) when a voltage step is applied adiabatically to a dielectric sample. It has a high theoretical efficiency (70%), higher than solid-state analogs like thermoelectrics (20%) or even the standard gas cooling cycle (50%) [1]. First theorized in 1878 by William Thomson [2], the high temperature changes (ΔT = 12 K) calculated for ferroelectric thin films [3], and the discovery of an anomalous electrocaloric effect in antiferroelectrics [3] have renewed its interest, with an eye put on its potential application as a solid-state cooling solution in integrated circuits.

Ferroelectric (FE) materials display what is regarded as the "conventional" electrocaloric effect (or positive electrocaloric effect), whereby the material increases temperature (ΔT > 0) when a voltage step is applied, and decreases temperature (ΔT < 0) when it is removed. In contrast, antiferroelectrics (AFE) display the opposite response: they decrease temperature (ΔT < 0) when an electric field $E$ is applied, and increase (ΔT > 0) when it is removed. The ability of antiferroelectrics to cool down despite electrostatic energy being pumped into them is intriguing, and different underlying mechanisms have been proposed for the negative electrocaloric effect [4], [5]. Recent experimental evidence indicates that, in the archetypal antiferroelectric $PbZrO_3$ (PZO), the so-called "giant" negative electrocaloric effect is due to the latent heat absorbed during the adiabatic field-induced AFE-FE transition, which is endothermic [6].

The direct link between the anomalous electrocaloric effect and antiferroelecric-ferroelectric switching implies that the field-induced nucleation and motion of the AFE-FE phase boundary will determine the dynamics of the large negative ECE, and ultimately the response speed of electrocaloric devices based on antiferroelectrics. In ferroelectrics, the study of domain wall dynamics [7]–[16] has been examined in detail on account of their relevance for ferroelectric memories. In contrast, there are far fewer works regarding the dynamics of the ferroelectric-paraelectric phase boundaries in FE [17], [18] or the antiferroelectric-ferroelectric ones in antiferroelectrics [19]–[23]. Yet, a priori, one cannot assume that the dynamics of domain walls will be the same as the dynamics phase boundaries: while the former separate different domains of within the same ferroelectric phase, the later separate

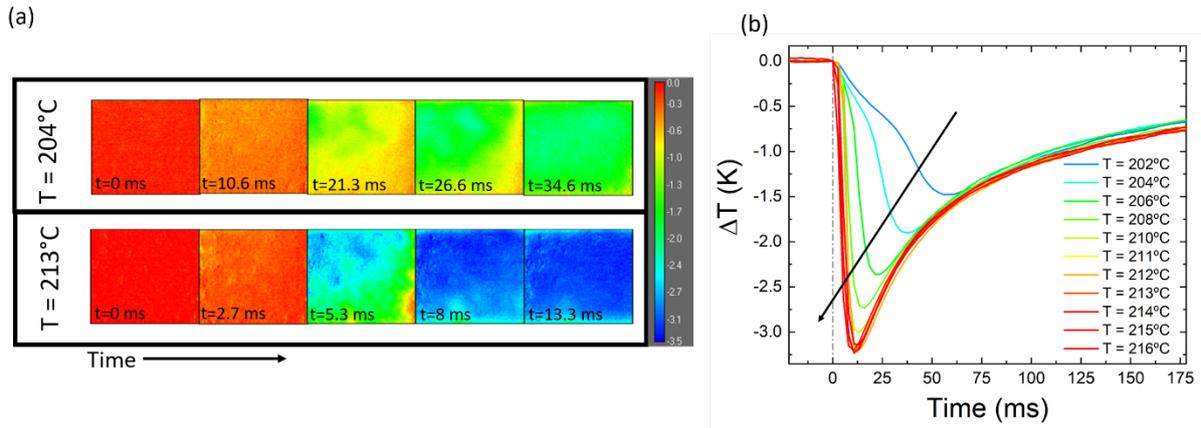

*Figure 1 (a) Antiferroelectric-ferroelectric switching dynamics observed by infrared imaging at two different temperatures for an electric field E = 35 kV/cm and sampling frequency of 376 Hz. Note that time labels for both temperatures differ. (b) Electrocaloric profile as a function of time. It can be clearly seen that, as the temperature increases, the switching time decreases.*

different phases –in antiferroelectrics, an antipolar phase from a field-induced polar one. It is the aim of this work to visualize and quantify the dynamics of the AFE-FE phase transition, determine its dominant mechanism (i.e nucleation-limited or propagation-limited), and measure the speed of the antiferroelectric phase-boundary.

The samples examined in this study are ceramics of pure $PbZrO_3$ (PZO), an archetypal antiferroelectric with a large negative electrocaloric effect [5], [24], [25]. Thanks to the electrocaloric temperature change concomitant with the AFE-FE phase transition, it is possible to use an infrared camera to observe how the electrocaloric front linked to the AFE switching nucleates/propagates across the sample in real time at a maximum frequency of 1253 Hz. In order to be able to switch the bulk ceramic capacitors with electric fields lower than the breakdown field, we work at temperatures close to, but below, the Curie temperature, which for PZO is $T_C$~230 °C. The electrocaloric response and switching dynamics are shown in Figure 1.

For a given electric field, the speed of the AFE-FE phase transition (and thus the speed of the negative ECE) increases with the sample's temperature (Figure 1). This increasing switching speed is a consequence of the higher thermal energy available to promote this transition, which weakens the antiferroelectric coupling and decreases the critical switching field as temperature approaches Tc. A direct consequence is that, for a given temperature, the ability of a field to switch the antiferroelectric depends on how much time the field is applied for. In Figure 1-b we see, for example, that while at 214 °C switching happens in 10 ms for the applied field of 35kV/cm, at 204 °C the same field needs to be applied for more than 25 ms for switching to happen. This behaviour is reminiscent of the frequency dependence of the coercive field in ferroelectric hysteresis [13].

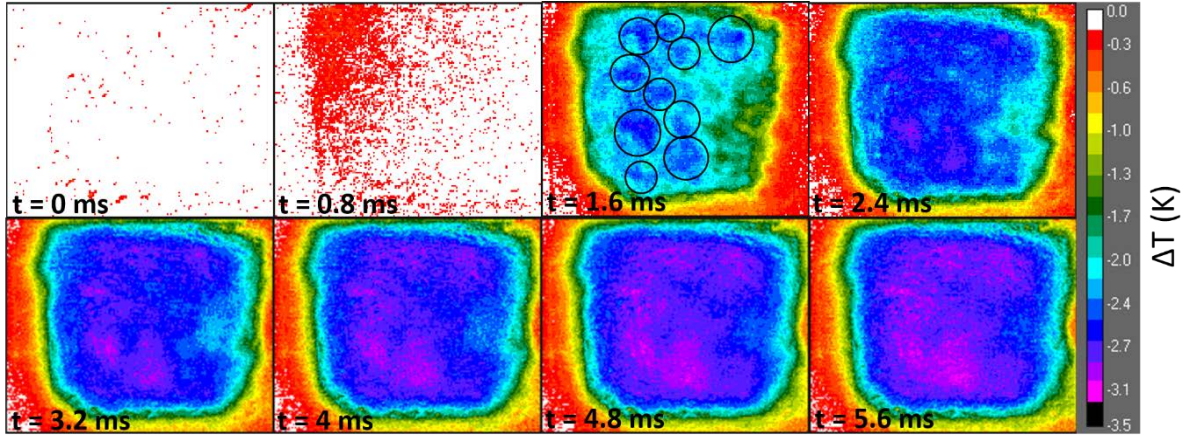

*Figure 2 Infrared imaging of the large negative electrocaloric effect at T = 216C and E = 28 kV/cm, acquired at a sampling rate of 1.25 kHz over a 1.15 mm² area. The black circles represent the main nucleation points. From these points the cold front propagates to cover the whole sample.*

For the fastest electrocaloric response (high field at high temperature), the switching is so fast that even at the fastest available frame-rate of 1250 fps most of the change happens within only a couple of frames (figure 2). Nevertheless, there is still enough information to quantify the switching speed and address the question of whether switching is dominated by nucleation or by sideways propagation. The IR pattern immediately after switching is consistent with multiple-site nucleation: we can observe a mottled seed pattern distributed across the sample (figure 2). The main nucleation centers (marked with black circles) can be distinguished as cold-spots (ΔT ~-2.5 K). From frame 2 to frame 3 in Figure 2, we can see that the whole sample already displays a negative ECE ($\overline{\Delta T}$ ~ -2.2C), which implies that the nucleation and sideways propagation process occurs in a time t << 800 µs while the complete switching of the sample occurs in $\Delta t_{EC}$ ~ 7.9 ms. Given that the maximum averaged negative electrocaloric effect is ΔT = -3.2 K, the electrocaloric response time $\tau_{EC} \equiv \frac{\Delta t_{EC}^{max}}{\Delta T_{max}} \approx 2.5 \frac{ms}{K}$.

The appearance of multiple nuclei across the sample makes the switching fast, but complicates the quantification of the phase boundary's sideways-propagation velocity given the limits of the infrared camera's acquisition speed. In order to obtain a single nucleation point, and thus be able to observe the boundary's movement in a propagation-limited process, we have modified the morphology of the sample to give it a wedged profile (see a scheme in Figure 3-b). In this way, the nucleation process starts at the thinnest side, where the electric field is more intense, and propagates sideways from there towards the thicker side. The sample was polished with a thickness gradient (Δd/Δx) of 16 µm/mm (75 µm difference across a 4.7 mm-long sample with an average thickness of 160 µm). The resulting switching process was filmed and the frames are shown in Figure 3. As expected, the phase transition starts at the thinner side of the wedge, and then sweeps across the sample like a cold "weather-front".

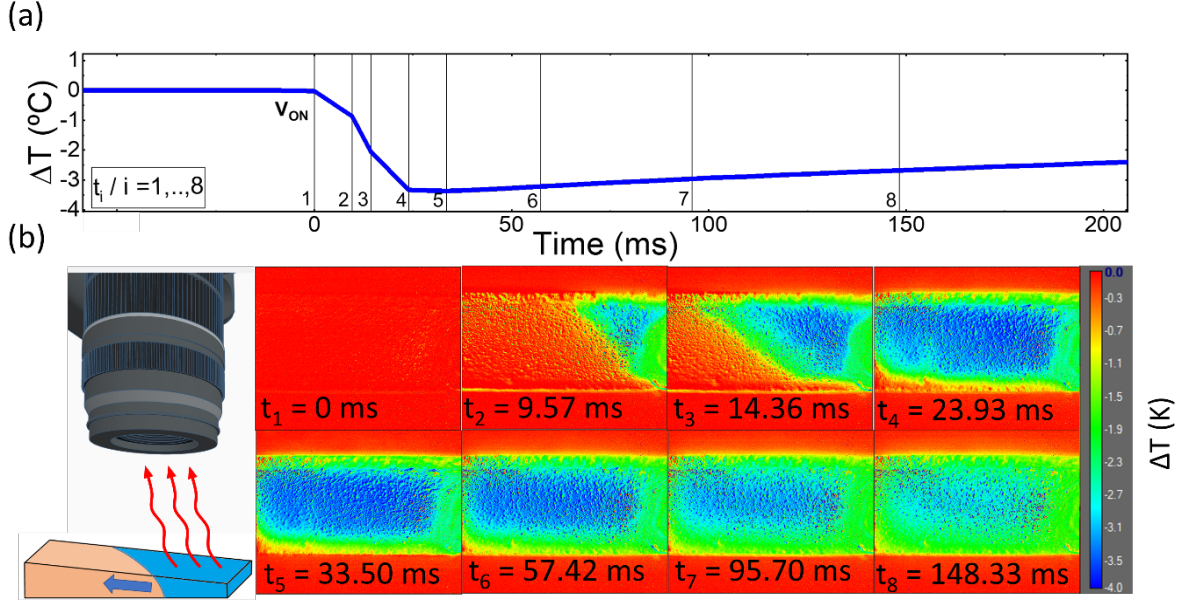

*Figure3 (a) Electrocaloric profile of PbZrO$_3$ at T = 216C when an average electric field E = 35 kV/cm is applied. (C) AFE-FE phase boundary propagating across the sample generating the negative ECE. The thickness gradient creates a sideways propagation of the phase boundary.*

For the largest applied field of 35 kV/cm, the front propagates sideways with a maximum velocity v = 20 cm/s, resulting in a device electrocaloric response time $\tau_{EC} = \frac{\Delta t_{EC}^{max}}{\Delta T_{max}} = 9.8 \frac{ms}{K}$. For E = 28 kV/cm, which is the same field used for the homogeneous-thickness sample in Figure 2, the electrocaloric response time ($\tau_{EC}$) is $\tau_{EC} = \frac{\Delta t_{EC}^{max}}{\Delta T_{max}} = 12 \frac{ms}{K}$. The response time of the propagation-limited dynamics (wedge) is therefore 4-5 times slower than that of the nucleation-limited parallel-plate capacitor.

The results demonstrate that thickness gradients can be used to modulate the dynamics of the electrocaloric effect. Thickness, however, is not the only magnitude that can be graded; gradients of temperature can also be used. For example, by sitting the sample at the edge of the heater, a temperature step can be introduced, thereby forcing the sample to bridge across two different points of its phase diagram. Since PZO has both an AFE-FE transition with negative electrocaloric effect and a ferroelectric-paraelectric one with a positive ECE (Figure 4), bridging the two transitions enables the possibility of a "Janus electrocaloric effect": a simultaneous positive and negative electrocaloric response upon applying a voltage to the sample (Figure 4).

The strong locking between antiferroelectric switching and electrocaloric response thus enables the observation in real time of the switching dynamics, and its manipulation by playing with the temperature and/or morphology of the sample. The switching in homogeneous antiferroelectric capacitors of ~150 μm is nucleation-limited, with "cold spots" appearing within less than 1ms and full device switch in ~7.9 ms with a concomitant ΔT = -3.2 ºC.

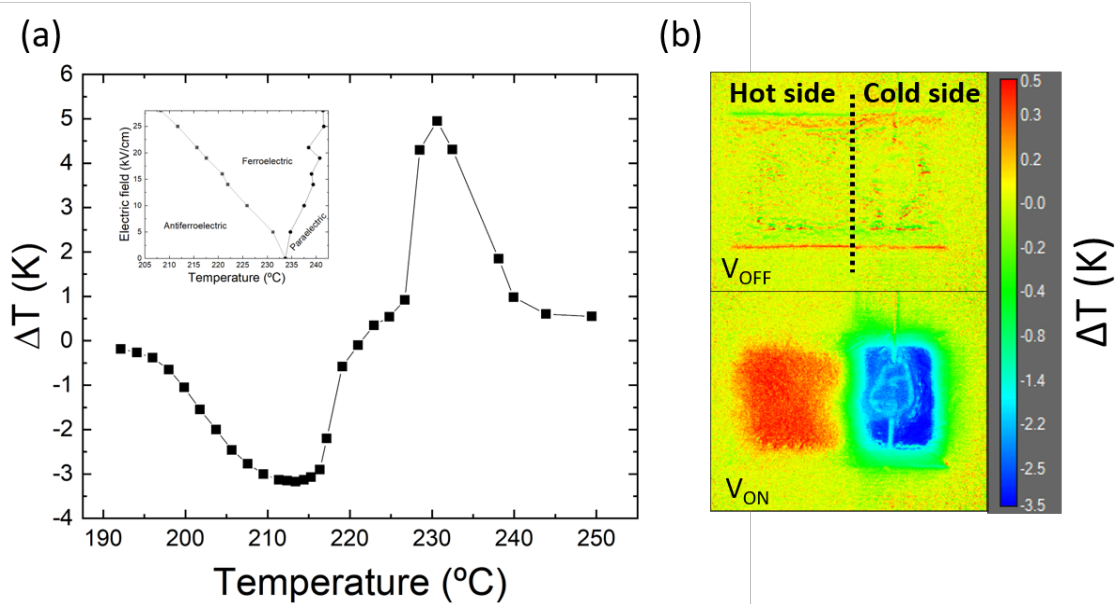

*Figure 4 (a) Electrocaloric profile of PbZrO$_3$ ceramics as a function of temperature at E = 35 kV/cm. Inset represents a simplified phase diagram modified from [6] (b) Realization of a Janus-like behavior whereby the sample displays simultaneously a negative and positive electrocaloric effect by setting both sides of the sample at different temperatures, namely T = 214°C and T= 223°C for the cold and hot sides, respectively. Note that the temperature bar represents a relative measure.*

In contrast, samples with thickness gradients can be used to concentrate the nucleation at one side of the sample and cause a sideways propagation of the thermal front, which may be useful for guiding the flow of heat (heat pumping or phononic guiding). Meanwhile, gradients of temperature can be used to bridge the phase diagram of PbZrO$_3$, so that application of voltage causes one side to get colder while the other gets warmer (Janus electrocaloric effect). The results emphasize the importance of antiferroelectric switching dynamics for electrocaloric device performance, as well as the exciting possibilities brought to bear by the combination of in-situ observation and gradient manipulation.

## Experimental methods

**Fabrication and sample preparation**

We have measured the electrocaloric effect of ceramics of the perovskite antiferroelectric archetype, PbZrO$_3$. The samples were made as described in [26] and disk-polished down to thicknesses between 100 and 150 microns with a Multiprep polishing system. 200 nm thick platinum electrodes where deposited by electron beam evaporation and platinum wires bonded with curated silver paste.

**Infrared characterization**

The electrocaloric dynamics was measured by infrared (IR) thermometry, using two different infrared cameras: a FLIR x6580sc and a FLIR SC5500. For the former, the acquisition speed used was 130 fps while, for the latter, it was 376 fps and 1253 fps. The lens was a G3 with a field of view (FOV) of 3.2 mm x 2.56 mm and 10 µm lateral resolution. Prior to the IR characterization, the samples were covered with an emissivity-calibrated black paint and located in a Linkam chamber (model T95-PE) coupled to a micropositioning system to allow their correct focusing, displacement, and setting its initial temperature. The electrocaloric effect was induced by a Keithley 2410 High Voltage Sourcemeter with voltage rise times of 300 microseconds. The measuring process is based on the dynamics of a Brayton cycle: applying a voltage step adiabatically, acquire the response and let the sample thermalize before adiabatically removing the field. The relative temperature changes acquired with the IR camera were measured with an accuracy of 0.1 K.

**ACKNOWLEDGEMENTS**

We acknowledge financial support to ICN2, which is funded by the CERCA programme/ Generalitat de Catalunya and by the Severo Ochoa programme of the Spanish Ministry of Economy, Industry and Competitiveness (MINECO, grant SEV-2017-0706). We also acknowledge support to Plan Nacional (MINECO Grant MAT2016-77100-C2-1-P and Grant BES-2016-077392), as well as the Agencia Estatal de Investigacion grant PID2019-108573GB-C21. R. Faye and E. Defay thank the Luxembourg National Research Fund (FNR) for funding part of this research through the projects CAMELHEAT/C17/MS/11703691/Defay. This work was also supported in part by the Spanish Ministry of Science, Innovation and Universities under the HIPERCELLS project (RTI2018-098392-B-I00), the Regional Government of the Generalitat de Catalunya under Grants 2017 SGR 1384 and 2017 SGR 00579. This work was also supported by the National Science Centre, Poland, within the project 2016/21/B/ST3/02242.